\documentclass[aps, prx, reprint, longbibliography, nofootinbib, floatfix]{revtex4-2}

\usepackage{slashed}
\usepackage{verbatim}
\usepackage[T1]{fontenc}
\usepackage[utf8]{inputenc}
\usepackage[colorlinks]{hyperref}
\usepackage{mathbbol}
\usepackage[dvipsnames]{xcolor}
\usepackage{graphicx}
\usepackage{orcidlink}
\usepackage{euscript}
\usepackage[normalem]{ulem}
\usepackage{svg}
\usepackage[english]{babel}
\usepackage{lipsum}
\usepackage{physics}
\usepackage{dcolumn}
\usepackage{tensor}
\usepackage{comment}
\usepackage{graphicx,color,overpic,mathtools}
\usepackage{amsthm,amsmath,amssymb,hyperref,mathrsfs}
\usepackage{braket,bm,bbm,setspace}
\usepackage{cancel}
\usepackage{float}
\hypersetup{
    colorlinks=false,
    pdfborder={0 0 0},
}
\usepackage{xargs}

\newcommand{\Amp}{\mathfrak{h}}

\definecolor{mypurple}{RGB}{130, 0, 130} 

 \hypersetup{
    colorlinks=true,
    linkcolor=mypurple, 
     citecolor=mypurple, 
    urlcolor=mypurple   
 }

\begin{document}

\title{Lensing and wave optics in the strong field of a black hole}

\author{Juno C. L. Chan~\textsuperscript{\orcidlink{0000-0002-3377-4737}}}
\author{Conor Dyson~\textsuperscript{\orcidlink{0000-0002-9742-9422} } }
\author{Matilde Garcia~\textsuperscript{\orcidlink{0009-0006-6899-1033} }}
\author{Jaime Redondo-Yuste~\textsuperscript{\orcidlink{0000-0003-3697-0319}}}
\author{Luka Vujeva~\textsuperscript{\orcidlink{0000-0001-7697-8361}}}
\affiliation{Center of Gravity, Niels Bohr Institute, Blegdamsvej 17, 2100 Copenhagen, Denmark}

\begin{abstract}

Gravitational waves (GWs) are lensed by matter, offering a unique probe of both the large-scale structure of the Universe and the fundamental properties of GW propagation. GWs can also be affected by wave optics effects when their wavelength is comparable to the size of the lens. While this regime has been well studied in the Newtonian approximation, the role of strong gravitational fields remains largely unexplored. This is particularly relevant for lensing by intermediate and supermassive black holes (BHs), which can occur near active galactic nuclei or in compact triple systems. In this work, we analyze the lensing of GWs by a non-rotating BH and compare our results to the Newtonian point-mass approximation. We construct frequency-dependent amplification factors that incorporate strong-field effects, revealing explicit polarization mixing and absorption by the event horizon. Using a fiducial GW event, we explore key phenomenological signatures of BH lensing, highlighting new observational opportunities to probe strong gravitational fields through GW lensing.
\end{abstract}

\maketitle

\section{Introduction}

Gravitational lensing offers a powerful tool to study wave propagation in curved spacetime. While extensively observed for electromagnetic waves, the lensing of gravitational waves (GWs) remains undetected. However, both upgraded current-generation detectors~\cite{KAGRA:2013rdx,LIGOScientific:2014pky,VIRGO:2014yos} and future observatories are expected to routinely detect lensed GW events \cite{Oguri:2019fix, Xu:2021bfn}. GWs undergo lensing in a similar way to light but with key differences: (i) their longer wavelengths allow us to probe novel gravitational lensing regimes \cite{Leung:2023lmq}, and (ii) their spin-2 nature leads to two polarization states, $+$ and $\times$, which can mix during propagation through strongly curved spacetimes, resulting in distinctive observational signatures ~\cite{Oancea:2023hgu,Oancea:2022szu,Andersson:2021,Harte:2022}.

The nature of GW lensing depends on both the properties of the waves and the lens. In this work, we focus on wave optics effects, which arise when the GW wavelength is comparable to the lens size. In this regime, a single lensed waveform (or image) is observed, but different frequencies experience varying magnifications and time delays \cite{Schneider:1992bmb}. These frequency-dependent distortions can significantly alter the waveform, posing challenges for identifying lensed GW events and accurately recovering source parameters \cite{Mishra:2023ddt,Chan:2024qmb}.

We investigate GW propagation through strong gravitational fields by considering compact lenses, specifically non-rotating black holes (BHs). Our interest lies in the upper left corner of Fig.~\ref{fig:lensing_reigmes}, corresponding to GW wavelengths comparable to the size of the lens, which in turn is also comparable to its mass. A key scenario where such effects arise is in triple systems: if a compact binary merges near a more massive BH companion, the emitted GWs will encode wave optics effects from the strong gravitational field of the BH~\cite{Cardoso:2021vjq}. Such configurations are expected to be common in dense globular clusters. Similarly, binary BH mergers near migration traps in active galactic nuclei could be lensed by a central supermassive BH, potentially exhibiting wave optics effects if the merging binary is sufficiently massive.

Wave optics effects in GW lensing have been extensively studied under the assumption that the lens produces only a weak gravitational potential. This approximation is valid for large-scale structures such as dark matter halos, galaxies, and galaxy clusters \cite{Grespan:2023cpa}. However, it becomes less accurate when the lens is a black hole (BH), where strong-field effects play a crucial role. In the Newtonian regime, an analytic solution exists under the point-mass lens (PL) approximation \cite{deguchi1986diffraction,Schneider:1992bmb}, which is particularly useful to study GW lensing~\cite{Nakamura:1997sw,Nakamura:2000ez,Takahashi:2003ix,Matsunaga_2006,Cao:2014oaa,Christian:2018vsi,Diego:2019lcd,Liao:2020hnx,Cheung:2020okf,Cremonese:2021puh,Yeung:2021chy,Wright:2021cbn,Urrutia:2021qak,Mishra:2021abc,Mishra:2021xzz,Bondarescu:2022srx,Caliskan:2022hbu,Caliskan:2023zqm, Mishra:2023ddt,Chan:2024qmb,Nerin:2024onm,Deka:2024ecp,Chen:2024xal,Villarrubia-Rojo:2024xcj}. However, this approximation neglects the strong gravitational potential near a BH and finite-size effects. Recent studies have taken steps toward addressing this limitation: numerical analyses of GW lensing by a nonrotating BH have revealed interesting magnification effects \cite{Yin:2023kzr}, while deep wave optics analyses have explored the problem in the long-wavelength regime \cite{Pijnenburg:2024btj}. Later studies have extended this to the case of spinning black holes~\cite{Motohashi:2021zyv, Kubota:2024zkv}.

The scattering of waves in BH spacetimes has long been an active area of research \cite{Futterman:1988ni}, with foundational theoretical work dating back to the 1970s \cite{Vishveshwara:1970zz, Press:1971wr, Teukolsky:1973ha, Teukolsky:1974yv, Virbhadra:1999nm, Virbhadra:2002ju}. In the geometric optics regime—complementary to our analysis—GW lensing is studied using unbound null geodesics in BH spacetimes \cite{Bozza:2001xd, Bozza:2002af, Bozza:2002zj}. In the wave optics regime, the propagation of each spherical harmonic mode of a GW can be formulated as a Schrödinger-like equation \cite{Regge:1957td, Zerilli:1970se}. The main challenge lies in decomposing an incoming GW, which is well approximated by a plane wave, into spherical harmonics and then resuming these components in a convergent manner.

In this work, we investigate GW lensing in the wave optics regime, comparing the PL approximation to the full strong-field regime. Our results extend those of \cite{Pijnenburg:2024btj} by going beyond the deep wave optics limit and improve upon \cite{Yin:2023kzr} by introducing a more efficient formalism that allows rapid evaluation of lensed GW signals for arbitrary source-lens-observer configurations. We construct an amplification factor that incorporates strong-field effects and explicitly accounts for the mixing between the $+$ and $\times$ GW polarizations. Finally, we demonstrate that while the PL approximation captures some qualitative features of BH-lensed GWs, significant differences arise due to polarization mixing and BH-induced absorption. 

This paper is structured as follows. In Sec.~\ref{sec:setup} we review GW lensing in the weak field regime, and the PL approximation. In Sec.~\ref{sec:lensing_to_perturb} we examine GW lensing in the strong field of a nonrotating BH. Our main results are presented in Sec.~\ref{sec:results}, followed by a discussion of future research directions in Sec.~\ref{sec:discussion}. Throughout this work, we adopt the Planck 2018 cosmology \cite{Planck:2018vyg}, and unless otherwise specified, we work with geometric units, $G=1=c$.

\section{GW lensing in weak fields}
\label{sec:lensing_to_perturb}

In this section, we review lensing of GWs in the weak field regime, placing special focus in the point mass lens model. We also comment on the main differences between wave and geometric optics, within the weak field regime.

\begin{figure}[t!]
    \centering
    \includegraphics[width=\linewidth]{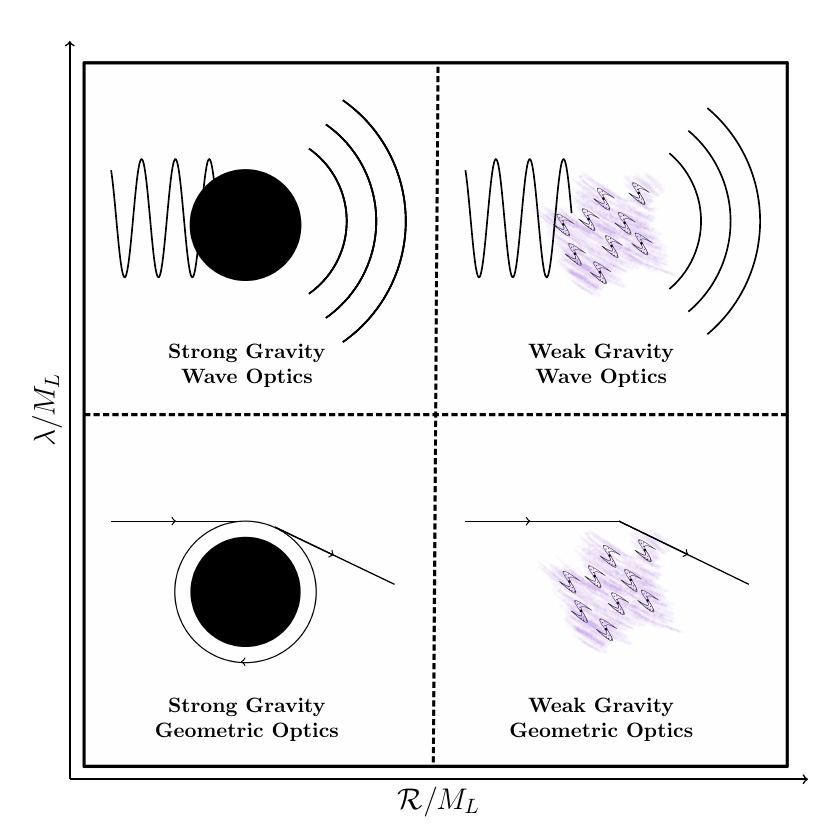}
    \caption{Diagram exemplifying the distinct regimes of GW lensing. We denote $M_L$ the lens mass, $\mathcal{R}$, its characteristic size and, $\lambda$ the wavelength of the incoming wave. In this work we are primarily concerned with the strong gravity, wave optics regime (topleft), particularly in the application of BH perturbation theory to explore the phenomenology of GW distortions in this part of parameter space. We will also make an explicit connection to lensing characteristics in the weak gravity, wave optics regime (topright).}
    \label{fig:lensing_reigmes}
\end{figure}

A gravitational potential $\Phi$ is considered to be a ``weak field'' when $|\Phi|<<1$. 
This applies to most astrophysical gravitational lenses such as stars, galaxies, and groups or clusters of galaxies, and allows one to treat their effects simply as a Newtonian potential, slightly distorting an otherwise flat geometry. Under this assumption, the gravitational wave is also treated under scalar wave theory, i.e., the changes in polarization are negligible.

Moreover, it is customary to consider the lens to be thin, relative to the characteristic distances of the problem. This allows us to approximate the lens potential $\Phi = \delta(r-r_l) \phi(\vec{x})$, $\vec{x}$ being the image position on the lens plane, and $r_l$ the distance from the observer to the lens \cite{Suyama_2005}. The propagation of the GWs is only affected by a two-dimensional projection $\phi(\vec{x})$ of the potential.

\begin{figure}[t!]
    \centering
    \includegraphics[width=\linewidth]{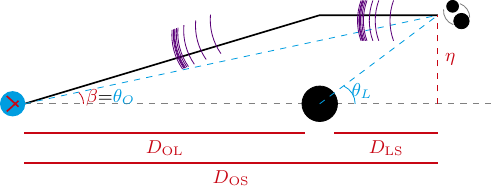}
    \caption{
    Schematic representation of the lensing configuration, with the angles and distances of in the language of the weak field approximation in red, and the strong field in blue. Note that the distances with capital letters are angular diameter distances (i.e. $D_{\rm OS} \neq D_{\rm OL} + D_{\rm LS}$, see Appendix~\ref{App:distances}).
    \label{fig:Cartoon}
    }
\end{figure}

Let $M \equiv M_L$ denote the mass of the lens, and $D_{\rm LS}, D_{\rm OL}$, and $D_{\rm OS}$ be the angular diameter distances between the lens and the source, between the observer and the lens, and between the observer and the source, respectively (see Fig.~\ref{fig:Cartoon}). Building upon these quantities, we can define a dimensionless angle $\theta_E$, commonly referred to as the Einstein radius,
\begin{equation}\label{eq:einrad}
    \theta_{E}  = \sqrt{\frac{4M D_{\rm LS}}{D_{\rm OL} D_{\rm OS}}} \, .
\end{equation}
This is nothing but the deflection angle by a PL. 

We also introduce the dimensionless source location in the lens plane $\vec{y} = \vec{\eta}/(\theta_E D_{\rm OS})$, where $\vec{\eta}$ is shown in Fig.~\ref{fig:Cartoon}. We refer the reader to Refs.~\cite{Takahashi:2003ix, Matsunaga_2006} for further clarification upon the variables commonly used in weak filed lensing. Finally, we can also construct a dimensionless frequency $\varpi$ at the lens, in terms of the angular frequency measured by the observer $\omega$, as
\begin{equation}\label{eq:dimensionless_w}
    \varpi = \frac{D_{\rm OL} D_{\rm OS}}{D_{\rm LS}}\theta_{\rm E}^2(1+z) \omega = 4M\omega(1+z)\, ,
\end{equation}
where $z$ is the lens redshift.
The modifications to the amplitude of a wave passing through the gravitational potential $\Phi$ are captured by a complex amplification factor, which, in terms of the dimensionless variables, and for a thin lens, is given by 
\begin{equation}\label{diffraction_integral}
    F(\varpi, \vec{y}) = \frac{\varpi}{2\pi i}\int d^2\vec{x} e^{i\varpi T_d(\vec{x},\vec{y})},
\end{equation}
where $T_d(\vec{x},\vec{y})$ is the dimensionless time delay surface given by
\begin{equation}\label{T_d}
    T_d(\vec{x},\vec{y}) = \frac{1}{2}|\vec{x} -\vec{y}|^2-\frac{\phi(\vec{x})}{ 2M} \, .
\end{equation}
We recall that the second term here is simply the projection onto the lens plane of the Newtonian potential of the lens.

We can distinguish the wave and geometric optics regimes from the diffraction integral, as represented in Fig.~\ref{fig:lensing_reigmes}. At high frequencies (small wavelengths), $\varpi \gg 1$, the integrand is highly oscillatory. Therefore the amplification factor is dominated by the stationary points of the time delay surface. The stationary phase approximation realizes this, by expanding the amplification factor as a sum over points $\vec{x}_i$, where  $\partial T_d/\partial x|_{\vec{x}=\vec{x}_i}=0$. Each of these stationary points corresponds to a distinct image location, each of them arriving with some time delay, and magnified or demagnified with respect to the original image. The possible existence of multiple images is a key aspect of the geometric optics regime. 

At lower frequencies, $\varpi \lesssim 1$, we transition to the wave optics regime (top right of Fig.~\ref{fig:lensing_reigmes}). In this case, the stationary phase approximation is no longer valid. For generic, complicated lens potentials the diffraction integral~\eqref{diffraction_integral} can be solved numerically~\cite{Villarrubia-Rojo:2024xcj, Yeung:2024pir, Cheung:2024ugg}. The case of a PL is, however, much simpler. Firstly, the potential is spherically symmetric, $\phi(\vec{x})=\phi(x)$, and thus the diffraction integral only depends on the one-dimensional coordinate $y\equiv|\vec{y}|$, as
\begin{equation}
    \begin{aligned}
        F(\varpi,y) =& -i\varpi e^{i\varpi y^2/2}\int_0^\infty dx x J_0(\varpi xy)\\
        &\times\exp\Bigl[i\varpi\Bigl(\frac{x^2}{2}-\frac{\phi(x)}{2M}\Bigr)\Bigr] \, .
    \end{aligned}
\end{equation}
Using the potential of a PL, which is given by $\phi(x) = 2M\ln|x|$\footnote{The projected potential is obtained through integration of the radial potential projected along the 2D lens plane. For a potential of the kind $\Phi(r) \propto -GM/r$ , the integral leads to a $\ln|x|$ projected potential--see full derivation in \cite{1999PThPS.133..137N}.}, the integral can be performed analytically, to obtain~\cite{Deguchi:1986zz,Nakamura:1997sw,Takahashi:2003ix}

\begin{align}
\label{pmaf}
    F(\varpi,y)=&\exp\bigg{[}\dfrac{\pi \varpi}{4}+i \dfrac{\varpi}{2} \bigg{\{}\ln\bigg{(}\dfrac{\varpi}{2}\bigg{)}-2\phi_m(y)\bigg{\}}\bigg{]} \nonumber \\
    &\times \Gamma \bigg{(}1- \dfrac{i}{2}\varpi\bigg{)} {_1}F_1\bigg{(}\dfrac{i}{2}\varpi, 1; \dfrac{i}{2}\varpi y^2\bigg{)} \, ,
\end{align}

where ${_1}F_1$ is the hypergeometric function~\cite{abramowitz1968handbook}, and $\phi_m(y)$ is a normalization constant dependent on $y$ such that the minimum value of the time delay is zero, and is given by
\begin{equation}
    \phi_m = \frac{1}{8}\Bigl(\sqrt{y^2+4}-y\Bigr)^2 - \ln\Bigl(\frac{y+\sqrt{y^2+4}}{2}\Bigr) \, .
\end{equation}
The lensed waveform is given simply by rescaling the frequency domain unlensed waveform with the amplification factor, $h^{\rm lensed}(f) = F(f) h^{\rm unlensed}(f)$, where $f$ here is the frequency observed at the detector. This is related to the dimensionless frequency $\varpi$ by $\varpi = 8\pi f M (1+z)$. Additionally, the impact parameter $y$ can equivalently be written in terms of the angle $\theta_L$ between the lens and the source, relative to the line of sight, via
\begin{equation}\label{weak-strong-map}
     y = \frac{\tan(\theta_L)}{\theta_{\rm E} }\frac{ D_{\rm LS}}{D_{\rm OL}} \, .
\end{equation}
This is particularly useful to connect this with the strong field description.

The description of GW lensing by a PL is particularly simple. However, much information is lost through the several approximations being made here. For instance, despite GWs being a spin $2$ wave, both polarizations are lensed equally. This is seen explicitly in Eq.~\eqref{pmaf}, since the amplification factor $F$ does not carry information about the helicity of the GWs. Additionally, the finite size of the BH is neglected, and as a consequence there is no way to account for absorption of GWs by the BH itself. All that Eq.~\eqref{pmaf} captures is the diffraction effects of a wave scattered by a pointlike potential. As we will see in following sections, the effects neglected here can be captured by treating the strong field of a BH accurately, at the expense of some technical difficulties.

\section{GW Lensing in the strong field}
\label{sec:setup}

We investigate the gravitational lensing of GWs by the strong field of a Schwarzschild BH. The source is a merging binary, located far enough from the lens that its emitted GWs can be approximated as plane waves, yet still satisfying $D_{\rm LS} /D_{\rm OS} \sim D_{\rm LS}/D_{\rm OL} \ll 1$. 

\subsection{GWs emitted by the source}

We first describe the GWs emitted by the source in its own reference frame. The source consists of two merging black holes, with a merger timescale short enough that the relative motion between the source and the lens can be neglected~\cite{Samsing:2024xlo, Samsing:2025rxq, Zwick:2025nbg}. In the transverse traceless gauge, the GWs emitted by the source can be written as 
\begin{equation}\label{eq:TTMetricSourceFrame}
    h_{\mu\nu}dx^\mu dx^\nu = \begin{pmatrix} 
    0 & 0 & 0 & 0 \\
    0 & h_+ & h_\times & 0\\
    0 & h_\times & -h_+ & 0 \\
    0 & 0 & 0 & 0 
    \end{pmatrix} \, , 
\end{equation}
where $h_{+(\times)}$ are the strain amplitudes of the $+$ ($\times$) polarizations, in Cartesian $(t,x,y,z)$ coordinates. In the frequency domain, the strain amplitudes take the form
\begin{equation}
    \begin{aligned}
        h_{+} =& \frac{e^{-i\omega(t-r)}}{2r}\Amp_{+}(f,\Omega_{\text{LS}} ) + \mathrm{c.c.} \, , \\
        h_{\times} =& \frac{ie^{-i\omega(t-r)}}{2r}\Amp_{\times}(f,\Omega_{\text{LS}}) + \mathrm{c.c.} \, ,
    \end{aligned}    
\end{equation}
where $\omega = 2\pi f(1+z)$ is the angular frequency at the source location, written in terms of the detector frequency $f$ measured at the observer, and $\Amp^{LM}_{+/\times}(f)$ are the (real valued) strain amplitudes for each polarization. The angles $\Omega_{\text{LS}} = \{\theta_{\text{LS}},\phi_{\text{LS}}\}$, describe the lens' position relative to the observer. The $+$ and $\times$ polarizations are not helicity eigenstates, as they lack a well defined spin weight. However helicity eigenstates can be obtained via
\begin{equation}
    h_{\pm2} = h_+ \pm i h_\times \, . 
\end{equation}
To analyze the GW behavior near the lens, we assume the lens is sufficiently distant from the source for the wave to be well approximated as a plane wave. Let $Z$ denote the line-of-sight distance measured from the source to a point $r$, and $r_{\rm LS}$ the physical distance between the source and the lens. We define $\theta_{OS},\phi_{OS}$ as the angular coordinates of the lens relative to the source. Following~\cite{Pijnenburg:2024btj} we approximate 
\begin{equation}
    \frac{e^{-i\omega(t-r)}}{r} = \frac{e^{-i\omega(t-r_{\rm LS})}}{r_{\rm LS}} e^{i\omega Z} \, .
\end{equation}
The helicity modes reaching the lens are then
\begin{equation}\label{eq:Helicity_Modes}
    h_{(\pm 2)} = H^{(\pm2)}e^{i\omega Z}+\bar{H}^{(\mp2)}e^{-i\omega Z} \, , 
\end{equation}
with 
\begin{equation}
    H^{\pm 2} = \frac{e^{-i\omega(t-r_{\rm LS})}}{2r_{\rm LS}}\Bigl(\Amp_+ \mp \Amp_\times\Bigr)\, .
\end{equation}
%

\subsection{BH scattering}

Once we have constructed the GWs impinging the lens, we need to solve the scattering problem. The lens is a Schwarzschild BH, with the metric given in the area gauge by
\begin{equation}\label{eq:Schwarzschild_metric}
    ds^2 = -fdt^2+f^{-1}dr^2+r^2d\Omega^2 \, ,  \quad f=1-\frac{2M}{r} \, , 
\end{equation}
with $M$ the BH mass, and $d\Omega^2$ the line element on the unit sphere. The linearized Einstein equations reduce to two wave equations for two master variables $\psi^{\bullet}_{\ell m}$ for each spherical harmonic constructed from the components of the metric perturbation, the Regge-Wheeler (RW) and Zerilli equations~\cite{Regge:1957td, Zerilli:1970se} 
\begin{equation}
    f\partial_r\Bigl(f\partial_r\psi_{\ell m}^{\bullet}\Bigr) + \Bigl(\omega^2 - V_{\ell m}^{\bullet}\Bigr)\psi_{\ell m}^{\bullet} = 0 \, , 
\end{equation}
where $\bullet = \{\mathrm{odd}, \mathrm{even}\}$. The respective potentials are given by 
\begin{equation}
    \begin{aligned}
        V_{\ell m}^{\rm odd} =& f\Bigl(\frac{\ell(\ell+1)}{r^2}-\frac{6M}{r^3}\Bigr) \, , \\
        V_{\ell m}^{\rm even} =& \frac{2f}{r^3} \frac{\lambda^2(1+\lambda)r^3+3\lambda^2Mr^2+9M^2(\lambda r+M)}{(3M+\lambda r)^2} \, ,
    \end{aligned}
\end{equation}
with $\lambda = (\ell+2)(\ell-1)/2$.
Despite their apparent differences, these potentials yield identical physics. Chandrasekhar~\cite{Chandrasekhar:1985kt} demonstrated this isospectrality by explicitly mapping the Zerilli equation to the RW equation through a particular change of variables~\footnote{The apparent complication can be seen as a consequence of the choice of master variables. An alternative choice of master variable in the even sector leads directly to the RW equation~\cite{Mukkamala:2024dxf}, albeit it comes with some drawbacks~\cite{Poisson:2025oic}.}.

At large distances ($r \gg M$), both equations reduce to the flat-space wave equation, leading to solutions of the form
\begin{equation}
    \psi^{\bullet}_{\ell m} \overset{r\to\infty}{\sim} A^{\rm in}_{\ell m}e^{-i\omega r_\star} + A^{\rm out}_{\ell m}e^{i\omega r_\star} \, , 
\end{equation}
where the wave is purely ingoing at the horizon, $\psi\sim e^{-i\omega r_\star}$ as $r\to 2M$ ($r_\star \to -\infty$). We define the reflectivity of the BH as 
\begin{equation}
    \mathcal{R}_{\ell m}(\omega) = \frac{A_{\ell m}^{\rm out}}{A_{\ell m}^{\rm in}} \equiv |\mathcal{R}_{\ell m}|e^{i\Theta_{\ell m}} \, .
\end{equation}
In the low-frequency regime ($\omega M \ll 1$), the waves are nearly perfectly reflected, $|\mathcal{R}_{\ell m}| = 1 + \mathscr{O}(M\omega)^{2\ell+1}$, with a phase shift given by~\cite{Poisson:1994yf}
\begin{equation}\label{eq:Low_Freq_Approx}
    \begin{aligned}
        \Theta_{\ell m} =& \pi(\ell+1)+2M\omega \Bigl[2\log(4M\omega)-\frac{(\ell-1)(\ell+3)}{\ell(\ell+1)}\Bigr]\\
        &-2M\omega\Bigl[H_\ell+H_{\ell-1}-2\gamma\Bigr]+\mathscr{O}(M\omega)^2 \, ,
    \end{aligned}
\end{equation}
where $H_n$ is the $n$-th harmonic number, and $\gamma\approx0.577$ the Euler-Mascheroni constant.

At high frequencies ($\omega M \gtrsim (\ell+1/2)/(3\sqrt{3})$) \cite{Cardoso:2008bp}, reflectivities vanish, as radiation with frequencies above this threshold--corresponding to the real part of the fundamental quasinormal mode of each harmonic--are absorbed. Thus, for $\omega M \gg 1$, only very large $\ell$ modes contribute, corresponding to null geodesics \cite{Ferrari:1984zz, Cardoso:2008bp}. We recover in this way the geometric optics limit.

\begin{figure}
    \centering
    \includegraphics[width=\linewidth]{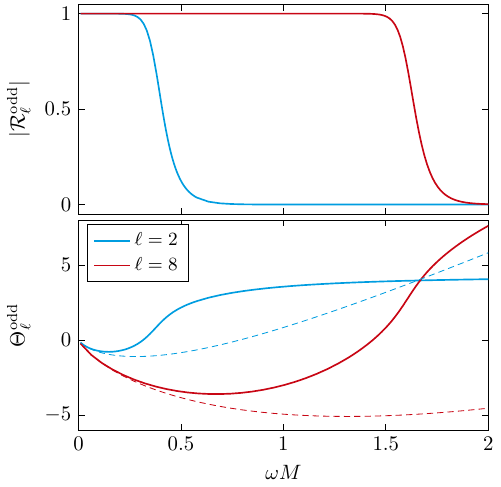}
    \caption{Reflectivity coefficient for the odd parity sector and the $\ell=2$ mode (in blue), and $\ell = 8$ (red), as a function of the GW frequency $\omega$ in units of the lens mass $M$. The upper panel shows the absolute value, whereas the lower panel shows the phase, where $\mathcal{R} = |\mathcal{R}|e^{i\Theta}$. In the bottom panel, the dashed lines show the phase obtained from the low frequency approximation~\eqref{eq:Low_Freq_Approx}. We observe that this is only a good approximation when $\omega M \ll 1$.
    \label{fig:Reflectivity_vs_w}}
\end{figure}

We implemented two methods to compute reflectivity coefficients across frequencies: one leveraging the \texttt{Black Hole Perturbation Toolkit} implementation of the Mano-Suzuki-Takasugi (MST) method~\cite{Mano:1996vt, BHPToolkit}, and another via direct RW equation integration. Both methods agree to subpercent accuracy up to $\ell = 30$, while the MST method enables reliable computations up to $\ell_{\rm max}=70$ for $\omega M \in[0.01,14]$. The reflectivity behavior is shown in Fig.~\ref{fig:Reflectivity_vs_w}. Notice that the reflectivity in the even parity sector is equal in magnitude to the odd sector (as required by isospectrality), and with a phase shift that can be computed through 
\begin{equation}
    \mathcal{R}^{\rm even}_{\ell m}(\omega) = \frac{(\ell+2)(\ell+1)\ell(\ell-1)+12iM\omega}{(\ell+2)(\ell+1)\ell(\ell-1)-12iM\omega}\mathcal{R}^{\rm odd}_{\ell m}(\omega) \, .
\end{equation}

The well-defined helicity modes of the GW strain $h_{(\pm 2)}$ can be decomposed in spin weighted spherical harmonics centered around the lens 
\begin{equation}\label{eq:Helicity_Decomposition}
    h_{(\pm 2)} = \sum_{\ell m} h^{(\pm 2)}_{\ell m}Y^{(\pm 2)}_{\ell m} \, .
\end{equation}
Each of these spherical modes is given in terms of the decomposition of the even and odd scalars by~\cite{Martel:2005ir}
\begin{equation}\label{eq:Psi_To_Helicity}
    rh^{(\pm 2)}_{\ell m} = \frac{1}{2}\sqrt{\frac{(\ell+2)!}{(\ell-2)!}}\Bigl(\psi^{\rm even}_{\ell m}\pm i\psi^{\rm odd}_{\ell m}\Bigr) \, .
\end{equation}
We now remind the reader of the expansion of a scalar plane wave into spherical harmonics, which is 
\begin{equation}
    e^{i\omega r\cos\theta} = \sum_{\ell}i^\ell \sqrt{2\ell+1}j_\ell(\omega r)P_\ell(\cos\theta) \, , 
\end{equation}
where $j_\ell(x)$ are spherical Bessel functions of the first kind, and $P_\ell(x)$ are Legendre polynomials. One can follow Refs.~\cite{Pitrou:2019ifq,Pijnenburg:2024btj} to  generalize the expansion to spin $2$ fields, finding, for the master variables
\begin{equation}\label{eq:Psi_From_Plane_Wave}
    \begin{aligned}
        \psi^{\rm even}_{\ell m=\pm s} =& \frac{iH^{\mp}}{\omega}\sqrt{\pi(2\ell+1)}\sqrt{\frac{(\ell-2)!}{(\ell+2)!}}\\
        &\quad\times \Bigl[(-1)^{\ell}e^{-i\omega r_\star}-e^{i\omega r_\star}\Bigr] + \mathrm{c.m.} \, , \\
        \psi^{\rm odd}_{\ell m}  =& \frac{im}{|m|}\psi^{\rm even}_{\ell m}  \, ,
    \end{aligned}
\end{equation}
where c.m. refers to the conjugate mode, obtained as $(-1)^m$ multiplied by the written contribution, after flipping the sign of $m$, sending $H^{\pm}\to H^{\mp}$, and taking the complex conjugate. We emphasize that only the modes with azimuthal number equal to the possible spin weights $|m| = s=2$ contribute. The reflectivities, however, do not depend on $m$ by virtue of spherical symmetry. 

This solution corresponds to the incoming spin-2 wave impinging the BH and passing through the system unaffected. We account for the strong field of the BH by rescaling the outgoing field with the BH reflectivity factors. In order to do so, we first subtract the incoming waveform, and then rescale the outgoing component as indicated below. We will later recover the ingoing piece, which leads to the component of the waveform that is directly propagated to the observer. Hence, we assume that the direct component does not interact with the potential of the lens. All the interaction is captured by the transmitted component, which is captured by

\begin{equation}
    \begin{aligned}
        \tilde{\psi}^{\rm even}_{\ell m=\pm s} =& \frac{iH^{\mp}}{\omega}\sqrt{\pi(2\ell+1)}\sqrt{\frac{(\ell-2)!}{(\ell+2)!}}\\
        &\quad\times \Bigl[1+(-1)^{\ell}\mathcal{R}^{\rm even}_{\ell m}\Bigr]e^{i\omega r_\star} + \mathrm{c.m.} \, ,
    \end{aligned}
\end{equation}
where the same relation as in Eq.~\eqref{eq:Psi_From_Plane_Wave} between even and odd parity modes holds. Inserting these into Eq.~\eqref{eq:Psi_To_Helicity}, and resumming them using Eq.~\eqref{eq:Helicity_Decomposition}, produces the desired lensed waveform. The nontrivial resummation procedure is discussed in more detail in Appendix~\ref{App:Convergence_Sum}, and we will return to this point later. Notice that, if we decompose each helicity mode as 
\begin{equation}\label{eq:Lensed_WF}
    r \tilde{h}_{(\pm 2)} = \tilde{H}^{(\pm 2)}e^{i\omega r_\star} + \bar{\tilde{H}}^{(\mp 2)} e^{-i\omega r_\star} \, , 
\end{equation}
we can easily write the lensed coefficients in terms of an amplification factor, as 
\begin{equation}\label{eq:Amp_Factors_General}
    \begin{aligned}
        \tilde{H}&^{(\pm 2)} = \mathcal{F}^\pm H^{(\mp 2)}/\omega \, , \\
        \mathcal{F}&^\pm (\Omega_{\rm OL}) = \frac{i\sqrt{\pi}}{2}\sum_{\ell,|m|=2} \sqrt{2\ell + 1} Y^{(\pm 2)}_{\ell m}(\Omega_{\rm OL}) f^\pm_{\ell m} \, , \\
        f&^\pm_{\ell m}= \Bigl[1+(-1)^{\ell}\mathcal{R}^{\rm even}_{\ell m}\Bigr]\mp\frac{m}{|m|}\Bigl[1+(-1)^{\ell}\mathcal{R}^{\rm odd}_{\ell m}\Bigr] \, ,
    \end{aligned}
\end{equation}
with $\Omega_{\rm OL}=\{\theta_{\rm OL},\phi_{\rm OL}\}$ the angular coordinates of the observer with respect to the lens.

    For each $\ell$ mode, there are two contributions, coming from $m = \pm s$, with $s = \pm 2$ in this case. At large $\ell$, the even and odd reflectivities become nearly identical, i.e., $\mathcal{R}_{\ell m}^{\rm even} - \mathcal{R}_{\ell m}^{\rm odd} \xrightarrow{\ell\to\infty} 0 $. As a result, the partial sums for $m=s$, which depend on the difference between these reflectivities, exhibit rapid convergence. In contrast, for $m = -s$ the partial sums do not converge. This is a known phenomenon in the multipolar expansion of the scattering of plane waves~\cite{Pijnenburg:2022pug,Pijnenburg:2024btj}, and convergence is only achieved within some finite radius of the BH for a given value of $\ell_{ \text{max} }$ \cite{Dyson:2024qrq}. Nevertheless, while the sum does not converge in the conventional sense, it possesses a well-defined Cesàro sum~\footnote{The Cesàro sum, in its basic form, is obtained by averaging the sequence of partial sums and taking the limit. More details are given in the Appendix~\ref{App:Convergence_Sum}.}, which we have verified numerically. In the Appendix~\ref{App:Convergence_Sum}, we provide extensive numerical evidence of the sum's effective convergence, as well as the errors introduced when truncating at finite $\ell$. We note that achieving convergence requires summing over a larger range of $\ell$ modes when the observer-lens angle $\theta_L$ is small. This imposes a practical limitation on probing smaller impact parameters, where richer strong-field effects will be observed.

\subsection{Waveform at the observer's location}

At the observer’s location, we write the total waveform as consisting of two contributions: (i) the lensed waveform, obtained by propagating~\eqref{eq:Lensed_WF} to the observer, and (ii) the direct, unlensed gravitational waves (GWs) from the source, which we previously subtracted to isolate the effects of the black hole's gravitational field. Using the notation introduced before, we express this as
\begin{equation}
    h^{\rm lensed}_{+/\times} = \tilde{h}_{+/\times} + h^{\rm unlensed}_{+/\times} \, .
\end{equation}
The lensed component takes the form
\begin{equation}
    \begin{aligned}
        \tilde{h}_{+} =& \frac{e^{-i\omega(t-r_{\rm LS}- r_{\rm OL})}}{ 4\omega r_{\rm LS}r_{\rm OL}}\Biggl[\Bigl(\mathcal{F}^++\mathcal{F}^-\Bigr)\Amp_+(\Omega_{\rm LS})\\
        &\hspace{2.25cm}+\Bigl(\mathcal{F}^+-\mathcal{F}^-\Bigl)\Amp_\times(\Omega_{\rm LS})\Biggr]+ \mathrm{c.c.} \, , \\
        \tilde{h}_{\times} =& \frac{e^{-i\omega(t-r_{\rm LS}- r_{\rm OL})}}{ 4i\omega r_{\rm LS}r_{\rm OL}}\Biggl[\Bigl(\mathcal{F}^+-\mathcal{F}^-\Bigr)\Amp_+(\Omega_{\rm LS})\\
        &\hspace{2.25cm}+\Bigl(\mathcal{F}^++\mathcal{F}^-\Bigr)\Amp_\times(\Omega_{\rm LS})\Biggr]+ \mathrm{c.c.}\, ,
    \end{aligned}
\end{equation}
where we see explicitly how the two polarization states mix due to the interaction with the black hole’s gravitational field. Notably, the GW amplitudes are evaluated at the sky location of the lens relative to the source $\Amp_{+/\times}(\Omega_{\rm LS})$, whereas the amplification factors depend explicitly on the angular coordinate of the observer relative to the lens, $\Omega_{\rm OL}$, via Eq.~\eqref{eq:Amp_Factors_General}.

The direct, unlensed component follows as
\begin{equation}
    \begin{aligned}
        h^{\rm unlensed}_{+} =& \frac{e^{-i\omega(t-r_{\rm OS})}}{2r_{\rm OS}}\Amp_{+}(\Omega_{\rm OS}) +\mathrm{c.c.}\, ,  \\
        h^{\rm unlensed}_{\times} =& \frac{ie^{-i\omega(t-r_{\rm OS})}}{2r_{\rm OS}}\Amp_{\times}(\Omega_{\rm OS}) +\mathrm{c.c.} \, ,  \\
    \end{aligned}
\end{equation}
where $r_{\rm OS}, \Omega_{\rm OS}$ are, respectively, the distance and the solid angle connecting the source and the observer's locations. 
We note that this formulation assumes no interaction between the direct component of the emitted gravitational wave and the potential of the lens, serving as the primary simplifying assumption of this work. A more sophisticated treatment of the direct piece must be required in order to recover the geometric optics results of the PL model in the high-frequency limit.
We focus on cases where the lens and source are relatively close compared to the observer, $r_{\rm LS}\ll r_{\rm OS} \sim r_{\rm OL}$. Under this approximation $\Omega_{\rm OS} \simeq \Omega_{\rm LS}$ (with equality holding in the case where source, lens, and observer are aligned). This allows us to express the total waveform compactly as
\begin{equation}
    h_{+/\times}^{\rm lensed} = \mathcal{F} h_{+/\times}^{\rm unlensed} + \mathcal{G} h_{\times/+}^{\rm unlensed} \, , 
\end{equation}
where $\mathcal{F}$ is the polarization preserving amplification factor
\begin{equation}
    \mathcal{F} = 1 + \frac{r_{\rm OS}e^{-i\omega(r_{\rm OS}-r_{\rm OL} - r_{\rm LS} )}}{2\omega r_{\rm OL}r_{\rm LS}} (\mathcal{F}^+ + \mathcal{F^-}) \, ,
\end{equation}
and $\mathcal{G}$ captures the polarization mixing contributions, defined as
\begin{equation}
    \mathcal{G} =  \frac{r_{\rm OS}e^{-i\omega(r_{\rm OS}-r_{\rm OL}-r_{\rm LS})}}{2\omega r_{\rm OL}r_{\rm LS}} (\mathcal{F}^+ - \mathcal{F^-}) \, .
\end{equation}
The exponential factor accounts for the geometric time delay between the lensed and the unlensed components. 

Restoring the dependence on the azimuthal angle $\phi_{\rm OL}$, we observe that both amplification factors, $\mathcal{F}$ and $\mathcal{G}$, can be expressed in terms of a single quantity $\mathbb{F}$, weighted differently depending on the azimuthal angle $\phi_{\rm OL}$, 
\begin{equation}
    \mathcal{F} = 1 + \mathbb{F} \cos(2\phi_{\rm OL}) \, , \qquad \mathcal{G} = i\mathbb{F} \sin(2\phi_{\rm OL}) \, .
\end{equation}
The total waveform is $h=h_+ \mathcal{A}_+ + h_\times \mathcal{A}_\times$, where $\mathcal{A}_{+/\times}$ are the detector's antenna response functions of the $+$ and $\times$ polarizations. Including the effects of lensing, this becomes
\begin{equation}\label{eq:polarization_mixing}
    \begin{aligned}
        h^{\rm lensed} =& \  h^{\rm unlensed} + \mathbb{F}\Bigl[\cos(2\phi_{\rm OL})h^{\rm unlensed} \\
        &+ i\sin(2\phi_{\rm OL})\Bigl(h_+^{\rm unlensed}\mathcal{A}_\times + h_\times^{\rm unlensed}\mathcal{A}_{+}\Bigr)\Bigr] \, .
    \end{aligned}
\end{equation}
The second term in the brackets explicitly represents polarization mixing, which arises purely from the geometric treatment of the gravitational field’s spin-2 nature, independent of the black hole’s strong-field effects. This contrasts with the PL case, where the GW propagates as a scalar wave. To highlight this effect, we set $\phi_{\rm OL}=\pi/6$, ensuring both terms in the brackets are comparable.

\begin{figure}
    \centering
    \includegraphics[width=\columnwidth]{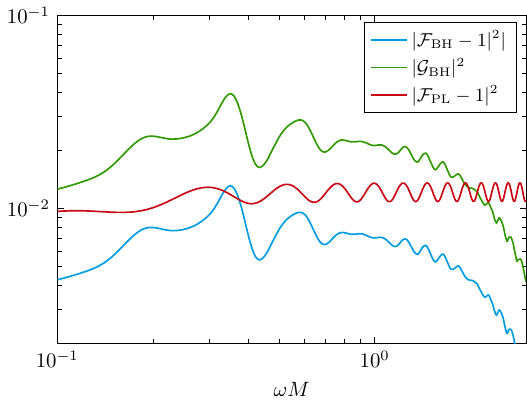}
    \caption{Absolute value of the amplification factor relative to the nonmagnified case, for the polarization preserving contribution, $|\mathcal{F}-1|^2$ (blue), and of the total polarization-mixing amplification $|\mathcal{G}|^2$ (green), as a function of the dimensionless frequency $\omega M$. The lens is located at $r_{\rm SL} =100M$, and the source has angular coordinates $(\theta_{\rm OS},\phi_{\rm OS}) = (\pi/6,\pi/6)$ with respect to the observer. In red, we give $|\mathcal{F}-1|^2$ for the case of a PL, corresponding to an impact parameter $y=2.69$. The dimensionless frequency $\varpi$ is related to $\omega M$ via Eq.~\eqref{eq:dimensionless_w}.}
    \label{fig:Amplification_Strong}
\end{figure}

We show the behavior of the amplification factor for both the polarization preserving, and the polarization mixing contributions in Fig.~\ref{fig:Amplification_Strong}. We choose a reference value of $r_{\rm LS} = 100M$, which is sufficiently far so that the plane wave approximation is valid, but small enough to clearly show the effects of the strong gravitational field of the lens. We find that the magnification of both polarization preserving and mixing contributions is greatest at intermediate frequencies, $\omega M \sim 10^{-1}-1$, quickly decaying at higher frequencies. This decay is associated to the absorption of GWs by the BH, which is not accounted for in the weak field regime. Additionally, we observe oscillatory features, which become prominent at approximately similar frequencies at which they appear for the PL approximation. These are smoking guns of diffraction and interference effects, which could be detectable with current detectors \cite{Lo:2024wqm}. We also recover some of the main features discussed in~\cite{Kubota:2024zkv}, albeit their amplification factor is defined at the level of the Weyl curvature scalar. In particular, we find that $|\mathcal{F}| \to 1$ as $\omega M \to 0$, and at high frequencies, we also encounter rapidly oscillatory features. 

In Fig.~\ref{fig:Amplification_Strong}, we additionally show the square of $\mathcal{F}-1$ computed using the PL approximation, for a situation matching the strong field calculation. We emphasize two main differences: (i) the strong field calculation outputs larger magnifications, relative to the PL approximation (for the polarization mixing contribution, due to our choice of the angle $\phi$), and (ii) at high frequencies, the amplification factor $|\mathcal{F}|\to 1$ and $\mathcal{G}\to 0$, since absorption by the BH becomes important. This effect is unaccounted for in the PL approximation. Nevertheless, it is remarkable that a simple approximation such as the PL can reproduce accurately several of the features observed in the complete, strong field calculation.

\section{Waveform comparison}
\label{sec:results}

Once we have constructed the amplification factors, we can directly compute the lensed waveform and analyze its most prominent features. Our amplification factor can be evaluated efficiently across a range of $\omega M$, and thus we capture key lensing effects in the strong gravitational regime throughout the entire inspiral merger ringdown. We will examine both the strong-field regime and the weak-field, PL approximation. For clarity, we will focus on a specific gravitational waveform resulting from the merger of spinless, equal mass BHs. This allows us to isolate the main contributions from strong-field lensing compared to the weak-field approximation. We will explore the lensed waveform in both the frequency and time domains.

\subsection{Lensed binary black hole system and wave optics effects}

To illustrate the effects of lensing on the GW signal, we simulate a GW waveform from the merger of two nonspinning BHs with equal masses (in the detector frame) $30~M_{\odot}$, a representative case for current ground-based detectors. We adopt a typical redshift $z=0.1$, which lies within the redshift horizon of GW detectors in the third LIGO-Virgo-KAGRA (LVK) observing run~\cite{KAGRA:2021vkt}, corresponding to a luminosity distance of $475.8~\rm{Mpc}$.
Inclination $\iota$ is set to zero, resulting in equal magnitude of the $+$ and $\times$ polarizations. We use the \texttt{IMRPhenomXPHM}~\cite{Pratten:2020ceb} waveform approximant in \texttt{LALSimulation}~\cite{lalsuite}.
A detailed summary of the waveform parameters is listed in Table~\ref{lalsim}, with definitions following~\cite{lalsuite}.

\begin{table}[h!]
\begin{tabular}{c c}
\hline
\hline
Properties &  \\
\hline
Primary mass $m_1$ & $30 M_\odot$\\
Secondary mass $m_2$ & $30 M_\odot$\\
Dimensionless spins & $0$\\
Redshift & $0.1$ \\
Luminosity distance & $475.8~\rm{Mpc}$\\
Inclination $\iota$ & $0$ \\
Waveform & \texttt{IMRPhenomXPHM}  \\
\hline
\end{tabular}
\caption{Basic parameters of the GW signal considered.}
\label{lalsim} 
\end{table}

The lens parameters are chosen to ensure the GW signal remains in the wave optics regime for the BH system under study. With a peak frequency of $f_{\rm peak} \sim 200 \mathrm{Hz}$, a lens mass in the range $M \in [100, 600]M_\odot$ satisfies $2\pi f_{\rm peak} (1+z) M \sim 0.6 - 3.6$, placing us in the intermediate regime between deep wave optics and geometric optics—our primary region of interest. This motivates our choice of lens mass within $M \in [100,600]M_\odot$. Given the comparable masses of the lens and merging BHs, the system likely forms a triple. If $r_{\rm LS} \omega \gg 1$, the amplification factor approaches $\mathcal{F} \to 1$ and $\mathcal{G} \to 0$, making lensing effects negligible. To avoid this regime, we set $r_{\rm LS}=100M$, which, while close to the merging BHs, remains sufficient to justify the plane wave approximation. The scales considered in this problem are comparable to those of~\cite{Cardoso:2021vjq}.

One of the main limitations of the resummation procedure required for the strong-field calculation is the divergence of the sum at $\theta_{\rm OL}=0$. Furthermore, for small angles $\theta_{\rm OL} \ll 1$, achieving convergence necessitates pushing the calculation to higher values of $\ell_{\rm max}$. With our current methods and $\ell_{\rm max}=70$, the smallest angle we can accurately and confidently resolve is $\theta_{\rm OL} = \pi/6$ (see Appendix~\ref{App:Convergence_Sum} for convergence tests). This corresponds to a large impact parameter, $y = 2.69$, following Eq.~\eqref{weak-strong-map}. Our inability to probe smaller impact parameters is a key limitation of this work. However, this limitation is purely technical, stemming from the challenges of resumming the spherical harmonic series. Alternative approaches, such as solving the problem in the time domain~\cite{Yin:2023kzr}, avoid this issue but come at the cost of significantly higher computational expense and a loss of analytical control. We also fix the azimuthal angle to $\phi_{\rm OL} = \pi/6$ to capture contributions from both polarization-preserving and polarization-mixing terms (see Eq.~\eqref{eq:polarization_mixing}).

We emphasize that when relating distances, it is important to account for cosmic expansion by converting physical distances into angular diameter distances. Although this effect is not critical here due to the moderate redshift, $z = 0.1$, properly incorporating cosmological distances ensures that the strong-field lensing formalism remains valid for arbitrarily distant lenses and sources. Details of this conversion are provided in Appendix~\ref{App:distances}.

\subsection{Frequency domain waveform}

\begin{figure}
    \centering
    \includegraphics[width=\columnwidth]{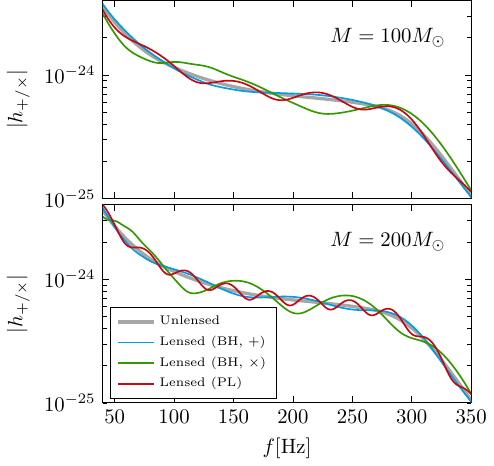}
    \caption{Frequency domain GW waveform for a lens mass $M = 100M_\odot$ (top), and $M=200M_\odot$ (bottom). The unlensed waveform is shown as a gray wide line, and it has equal magnitude for both $+$ and $\times$ polarizations. Under the PL approximation, the polarization content is not affected. The lensed waveform is shown as a red line. Finally, when accounting for the strong field effects, both polarizations are affected differently, as shown in blue ($+$), and green ($\times$).}
    \label{fig:FD_waveform}
\end{figure}

We begin by analyzing the lensed waveform in the frequency domain, shown in Fig.~\ref{fig:FD_waveform}. As explained before, when accounting for the strong field effects, the two polarization modes, $+$ and $\times$ are lensed differently. This effect, which is a direct consequence of the tensorial nature of GWs, is unaccounted for in the PL approximation. Despite having a large impact parameter, we observe noticeable lensing-induced distortions to the GW. These are more evident for the strong field, BH lens
, $M = 100M_\odot$ (upper panel). Qualitatively, the PL approximation seems to capture the most prominent features of the lensed waveform. 

In the high mass case, $M = 200M_\odot$ (lower panel), we observe that while the PL approximation predicts a highly oscillatory strain at all frequencies, the strong field calculation shows the amplitude of these oscillations decreasing with increasing frequency. This is the same feature shown in Fig.~\ref{fig:Amplification_Strong} at high frequencies, GWs of the lensed contribution are absorbed by the BH, and we recover only the unlensed component. 

\begin{figure}[t!]
    \centering
    \includegraphics[width=\columnwidth]{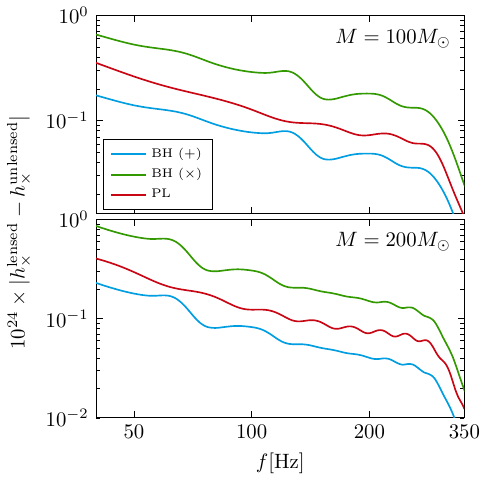}
    \caption{Absolute value of the difference between the lensed and unlensed waveforms when lensed by a BH taking into account the effects of the strong gravitational field, for the $+$ (blue) and $\times$ (green), and using the point mass lens approximation (red). Recall that the PL approximation lenses equally both polarizations. The top (bottom) panel shows the case where the lens mass is $M=100M_\odot$ (respectively, $M=200M_\odot$).}
    \label{fig:FD_delta}
\end{figure}

We examine the differences in the lensed and unlensed waveforms in Fig.~\ref{fig:FD_delta}. The figure shows the rescaled difference between lensed and unlensed contributions, for the two lens masses considered, and the different polarizations. For the lens mass $M=100M_\odot$, the peak frequency satisfies $\omega_{\rm peak} M \approx 1$, placing the system primarily in the wave optics regime. In this regime, GW absorption by the BH is less significant, and the PL approximation captures some of the qualitative features of the lensed waveform. At higher frequencies, however, we observe modulations that likely signal the onset of lensing effects probing the BH's potential structure. Notably, relating these modulations to the BH's fundamental frequencies could provide a novel approach for extracting fundamental physics from GW lensing \cite{Ferrari:1984zz, Cardoso:2008bp, Cardoso:2021vjq}. Although the PL approximation serves as a good first-order approximation, it fails to capture both the quantitative details and the polarization differences associated with strong-field lensing effects.

In the bottom panel, we consider the lens mass $M=200M_\odot$, which begins to probe the regime where $\omega M \gtrsim 1$. Here, BH absorption becomes more significant, and higher frequencies are not lensed in our strong-field calculation. This is partly due to an assumption in our model that the transmitted component is unaffected by the gravitational field of the BH—a reasonable assumption in the wave optics regime, but less valid as the system approaches the geometric optics regime. In the latter, GWs propagate along null geodesics of the metric. Further work is needed to establish a robust connection between the wave optics and geometric optics regimes in the context of GW lensing by BHs.

Overall, we find that gravitational lensing can lead to up to $\sim 10\%$ differences in the frequency domain strain. This result is for a particularly large impact parameter, and it is expected that more significant modifications to the waveform will emerge at smaller impact parameters. Consequently, we do not observe the large magnifications reported in Ref.~\cite{Yin:2023kzr}, which was based on the perfectly aligned scenario with an observer located in the strong field.

\begin{figure*}
    \includegraphics[width=\textwidth]{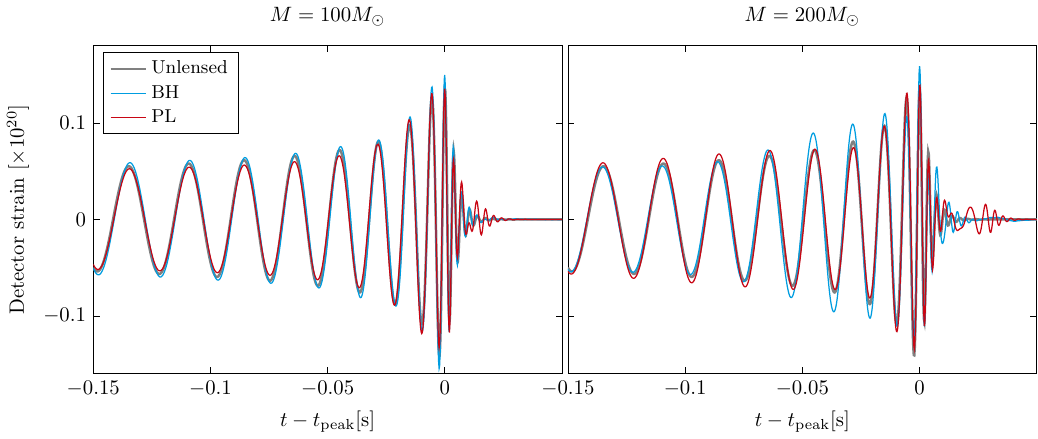}
    \caption{Time domain GW waveform observed at the detector aligned at peak, unlensed (black), lensed by a BH, accounting for strong field effects (blue), and in the weak field, PL approximation (red). The left (respectively, right) panel corresponds to a lens mass $M=100M_\odot$ (resp. $M=200M_\odot$). In order to obtain the detector strain, we assume a sky location given by the right ascension to be $\pi/8$, declination to be $\pi/4$ and polarization angle to be $\pi/4$.
    \label{fig:Time_Domain}
    }
\end{figure*}

\subsection{Time domain waveform}

While the effects of lensing are evident in the frequency-domain waveform, it is instructive to compare the waveforms in the time domain. For clarity, we fix the source’s sky location to right ascension to be $\pi/8$, declination to be $\pi/4$ and polarization angle to be $\pi/4$, ensuring comparable detector pattern functions, $\mathcal{A}_+ \sim \mathcal{A}_\times$. The measured strain at the detector is given by $h = \mathcal{A}_+ h_+ + \mathcal{A}_\times h_\times$, following the definition in~\cite{Anderson:2000yy}.

Figure \ref{fig:Time_Domain} presents the time-domain waveforms
for both lens masses previously considered. We display three cases—unlensed, lensed by a BH in the strong-field regime, and lensed using the weak-field PL approximation—aligning them by their peak amplitudes. Due to the large impact parameter, the lensed and unlensed signals exhibit a very small dephasing. However, magnification effects are apparent. For the low mass case, $M=100M_\odot$, the PL approximation predicts a slight demagnification, while the strong field calculation reveals a subtle magnification, which becomes most relevant close to the merger. The situation is different for the high mass scenario, $M=200M_\odot$. Both methods show magnification, which is stronger in the BH scenario. Notably, at higher frequencies, black hole absorption reduces distortions near the merger in the strong-field calculation—an effect absent in the PL approximation. On the other hand, the distortions to the waveform near the merger are very relevant for the PL approximation. We also observe the presence of the second image after the merger in both the PL approximation, with an increasing relative time delay and a similar relative magnification ($\sim 0.1$), indicating the contribution of geometrical optics limit in this regime.

\begin{figure*}[t!]
    \centering
    \includegraphics[width=\textwidth]{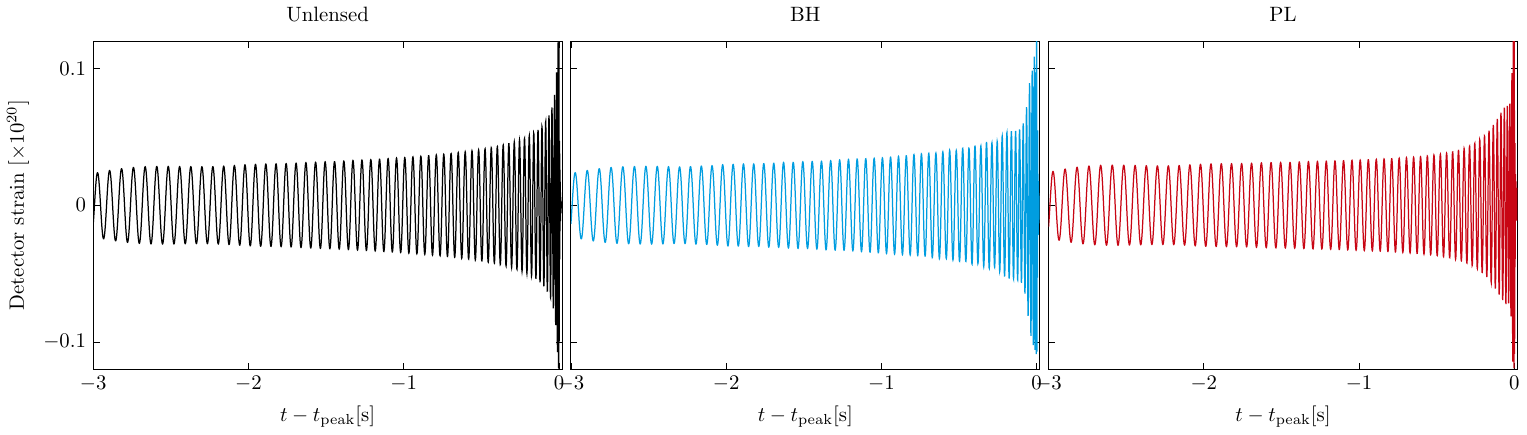}
    \caption{GW strain at the detector for the original waveform (black, left), the waveform lensed by a BH with mass $M=200M_\odot$ (blue, middle), and lensed by a PL with the same mass, in the weak field regime (red, right). 
    Small modulations of the waveform appear in this last case.
    }
    \label{fig:Inspirals}
\end{figure*}

We further analyze distortions during the inspiral phase for the more massive lens. Figure \ref{fig:Inspirals} focuses on the early inspiral for the unlensed, BH-lensed, and PL-lensed waveforms. Despite the large impact parameter, the PL approximation exhibits a subtle beating pattern, arising from the convolution of the waveform with the high-frequency oscillations of the amplification factor (see Fig.~\ref{fig:Amplification_Strong}). In contrast, the strong-field amplification factor rapidly approaches $|\mathcal{F}|\to 1$ as $\omega M \gg 1$, suppressing such interference effects. For smaller impact parameters, we expect more pronounced beating patterns in both regimes~\cite{Takahashi:2003ix,Cardoso:2021vjq}. These patterns encode information about the absorption of high-frequency radiation by the black hole, and resolving their structure will be the focus of future work.

\subsection{Mismatch}

We quantify the differences in the results between taking the full strong field nature of the BH into account as opposed to making the point mass lens approximation. 
By changing the mass of the lens $M \in [100,600]M_\odot$ while keeping the GW being lensed fixed, we can scan different regimes--from the pure wave optics regime to the transition toward geometric optics.
Small modulations of the waveform appear in this last case.
In order to quantify the distortions introduced to the GW by gravitational lensing, we define the mismatch as
\begin{equation}\label{mismatch}
    \epsilon = 1 - \mathrm{max}\left(\frac{(h_1|h_2)}{\sqrt{(h_1|h_1),(h_2|h_2)}}\right)_{t_0,\varphi_0}\,,
\end{equation}

where the second term is the match maximized over global time delays  $t_0$ and phases $\varphi_0$ and $(a|b)$ is the noise-weighted inner product, given by
\begin{equation}\label{inner_product}
    (a|b) = 4\mathrm{Re}\int_0^{\infty}\frac{\tilde{a}^*(f)\tilde{b}(f)}{S_n(f)}df,
\end{equation}
$\tilde{h}(f)$ being the Fourier transforms of the time-domain signals $h(t)$ , asterisks denote complex conjugation, and $S_n(f)$ is the noise power spectral density of the third LVK observing run. The mismatch for different polarization modes is computed individually, except in the case of the point mass lens model.

\begin{figure}
    \centering
    \includegraphics[width=\columnwidth]{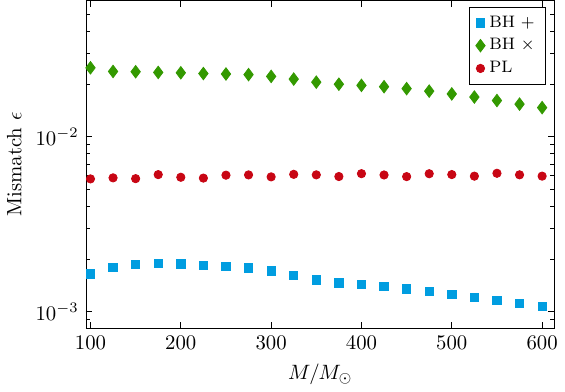}
    \caption{Mismatch $\epsilon$ between lensed and unlensed waveforms defined in Eq.~\eqref{mismatch} as a function of the lens mass, for the PL approximation (red), and for the BH lens ($+$ and $\times$ polarizations in blue, and green, respectively). }
    \label{fig:Mismatch}
\end{figure}

Figure \ref{fig:Mismatch} shows the mismatch between lensed and unlensed waveforms both for the BH and the point mass lenses, as a function of the lens mass. The mismatch is significantly different between both methods considered here. Although part of this difference is due to the polarization mixing contributions, the dependence of the mismatch on the lens mass is also different. We find that $\epsilon$ becomes smaller for larger values of the lens mass. This is due to a larger fraction of the lensed component of the waveform being absorbed by the BH. This is consistent with the findings of~\cite{Yin:2023kzr}, where they also report a better agreement between the lensed waveform and the reconstructed, best-fit waveform, for larger lens masses, compared to smaller lens masses. Due to the large impact parameter considered here, we only find a small mismatch, $\epsilon \lesssim 1\%$, whereas~\cite{Yin:2023kzr} considers the perfectly aligned case. However, the methods presented here can be improved upon in order to understand the dependence of the mismatch not only on the lens mass, but also on the impact parameter, in the strong field regime. 

\section{Discussion}
\label{sec:discussion}

In this work, we explored the gravitational lensing of GWs by the strong gravitational field of a Schwarzschild BH. While this problem has a long history, a direct comparison between wave scattering in BH perturbation theory and the weak-field approximation in GW lensing was lacking. We present, for the first time, a semianalytical study of GW lensing in the strong-field, wave-optics regime. Previous works had either focused on the deep wave-optics limit ($\omega M \ll 1$)~\cite{Pijnenburg:2024btj}, considered only wave scattering~\cite{Motohashi:2021zyv, Kubota:2024zkv}, or relied on fully numerical time-domain analyses~\cite{Yin:2023kzr}.

We began by reviewing the point mass lens approximation in the weak field regime, and compared it with the equivalent calculation in the strong field. This involves decomposing the incoming plane wave in spherical harmonics, propagating those through the curved geometry, where they are modified through the BH reflection coefficients, and finally reconstructing the plane wave by resumming the multipolar expansion. To ensure convergence of the resummation, we introduced an averaging procedure dubbed Cesàro summation. Despite this, convergence remains slow at smaller inclination angles, limiting our ability to explore small impact parameters. This could be improved by incorporating, e.g., semianalytical approximations for the reflectivity coefficients for large $\ell$ modes, though further exploration is needed. 
Using this resummation, we derived a direct expression for the GW amplification factor in the strong-field regime, parametrized by the BH's reflection coefficients. This framework is flexible and can be extended to study lensing by other compact objects, including neutron stars, exotic objects, or BHs in modified gravity theories. A restricting assumption of our work is that we decompose the GWs in a directly transmitted component, which is unaffected by the BH potential, and a lensed component which is distorted following~\cite{Pijnenburg:2024btj}. In the geometric optics limit, the lensed component is almost fully absorbed by the BH, and we only recovered the unlensed waveform. Relaxing this assumption will allow for a more clear connection between the wave and geometric optics limits, in the strong field regime.

We applied our formalism to a simple GW signal (not unlike the first observed LVK event), considering an intermediate-mass BH lens ($M = 100 - 600 M_\odot$) in the $\omega M \sim 1$ regime. The lensed waveform is magnified, with a frequency dependence that shows interference effects. For $\omega M \lesssim 1$ the point mass lens approximation captures accurately the main features of gravitational lensing by a BH. We find disagreements mostly stemming from the polarization mixing contributions, which are neglected in the weak field regime, but important when the lens is a compact object such as a BH. For $\omega M \gtrsim 1$, both methods disagree, as e.g. GW absorption by the BH becomes significant. This qualitative agreement is remarkable and should not be overlooked. It suggests that the point mass lens approximation could be systematically improved to account for absorption (or generic finite size effects) and polarization mixing, offering a practical alternative to full relativistic treatments. 

In the time domain, we identify significant magnification effects due to gravitational lensing, which are most important near merger. The fully relativistic lensed signal shows some differences with respect to the point lens approximation, which need to be scrutinized further. In particular, when the lens mass is relatively high, the point mass lens approximation predicts the appearance of beating patterns during the inspiral. These are absent for the strongly gravitating, BH lens--the lensed component is mostly absorbed by the BH at large values of $\omega M$. Due to the large impact parameter that we consider, we do not observe a significant phase shift. However, significant dephasing between lensed and unlensed waveforms is expected for smaller impact parameters (smaller values of the angle $\theta_S$), also in the strong field regime. Finally, we show the behavior of the mismatch between lensed and unlensed components as a function of the lens mass. While the magnitude of the mismatch does not vary significantly for the point mass lens case, oscillating around $\epsilon \sim 0.7\%$, the mismatch evolves significantly for the strongly gravitating lens. We find, consistent with the result of~\cite{Yin:2023kzr} that the mismatch decreases with lens mass.

Our results bridge the gap between prior strong-field lensing studies~\cite{Yin:2023kzr, Kubota:2024zkv, Pijnenburg:2024btj}, which either restricted the observer to be positioned in the vicinity of the lens or relied on low-frequency expansions. However, our framework cannot yet describe observers exactly aligned with the lens and source ($\theta_{\rm OL} = 0$), where maximum magnification is expected to occur. We also find no arrival-time difference between $|h_+|$ and $|h_\times|$, consistent with previous findings~\cite{Oancea:2022szu} for nonrotating BHs. This is not the case for a rotating BH lens.

This work opens multiple avenues for studying GW lensing using BH perturbation theory. A key challenge is extending the framework to spinning BHs, where ergoregions and broken spherical symmetry enable richer phenomenology~\cite{Motohashi:2021zyv,Kubota:2024zkv}. Further investigation of diverse source-lens configurations, beyond those discussed here, may reveal additional structures in GW lensing through strong gravitational fields. By leveraging techniques from BH perturbation theory, we advance the understanding of how GW lensing probes the strong-field regime of gravity.

\section*{Acknowledgements}
We thank Jose Ezquiaga, Vitor Cardoso, and Miguel Zumalacárregui for valuable discussions and insightful comments.
The Center of Gravity is a Center of Excellence funded by the
Danish National Research Foundation under grant No. 184.
This work makes use of the Black Hole Perturbation Toolkit~\cite{BHPToolkit}.
We thank the The Erwin Schrödinger International Institute for Mathematics and Physics (ESI) and
the organizers of the workshop ``Lensing and Wave Optics in
Strong Gravity''.
We acknowledge support by VILLUM Foundation (grant no. VIL37766 and no.~VIL53101) and the DNRF Chair program (grant no. DNRF162) by the Danish National Research Foundation.
The Tycho supercomputer hosted at the SCIENCE HPC center at the University of Copenhagen was used for supporting this work.  

\bibliography{biblio}

\clearpage 

\appendix 

\section{Convergence of the multipolar sum}
\label{App:Convergence_Sum}

Decomposing a plane wave into spherical harmonics, and then resumming it, leads to a multipolar sum~\cite{Pijnenburg:2022pug} that does not converge in the usual sense. The sum of our interest is the following
\begin{equation}
    \mathcal{F}^{\pm} \sim \sum_{\ell=2}^\infty \Bigl(f^\pm_{\ell 2}Y^{(\pm 2)}_{\ell 2} + f^\pm_{\ell -2}Y^{(\pm 2)}_{\ell -2}\Bigr) \, , \qquad s = \pm 2 \, .
\end{equation}
We can decompose this into two terms,
\begin{equation}
    \begin{aligned}
        \mathcal{F}^{\pm} =& \mathcal{F}^{\pm}_{m=s} + \mathcal{F}^{\pm}_{m=-s} \, , \\
        \mathcal{F}^{\pm}_{m=s} =& \sum_{\ell=2}^\infty f_{\ell \pm 2}Y^{(\pm 2)}_{\ell s} \, , \qquad \mathcal{F}^\pm_{m=-s} = \sum_{\ell=2}^\infty f_{\ell \mp 2}Y^{(\pm 2)}_{\ell \mp 2} \, .
    \end{aligned}
\end{equation}
Referring to the definition of the $f^\pm_{\ell m}$ in Eq.~\eqref{eq:Amp_Factors_General}, notice how the terms with $m=s$ involve the difference between the contributions of the odd and even sectors, whereas the $m=-s$ involve the sum of both contributions. It should not come as a surprise, then, that while the first kind of terms converges rapidly in the usual sense, the second kind of terms does not. 

\begin{figure}
    \centering
    \includegraphics[width=\columnwidth]{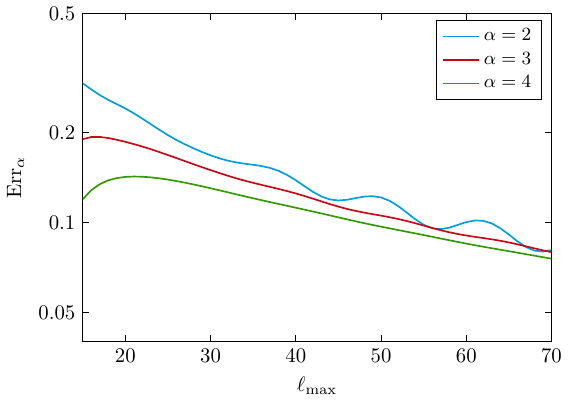}
    \caption{Relative error in the evaluation of $\mathcal{F}^+$ as a function of the $\ell_{\rm max}$ at which the sums are being truncated, for different values of the order of the Cesàro summation. Asymptotically, the error committed is at the percentage level.}
    \label{fig:Error_Sums}
\end{figure}

This is due to an oscillatory behavior. A way to ``cure'' this pathology, and give this sum a finite value, receives the name of Cesàro summation. In its classical sense, the Cesàro sum is the average of the sequence of partial sums. More generally, we can define the order-$\alpha$ Cesàro sum of a sequence $\{z_n\}$ as 
\begin{figure}[h!]
    \centering
    \includegraphics[width=\columnwidth]{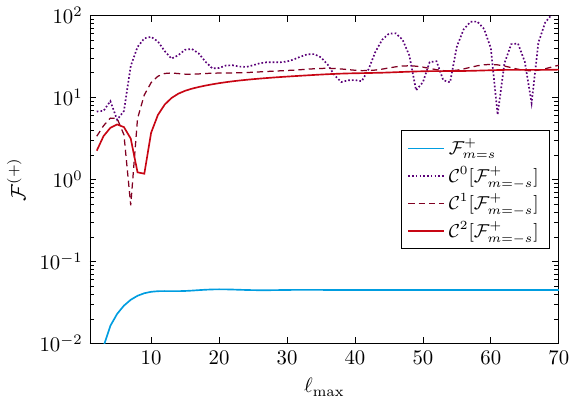}
    \caption{Plus (blue) and minus (red colors) contributions to the total amplification factor with positive spin weight $\mathcal{F}^+$, as a function of the highest angular harmonic included in the sum $\ell_{\rm max}$. Here we set $\theta_{\rm OL} =\pi/6,\phi_{\rm OL}=\pi/6,\omega M =0.5$. The contribution with $m=s$, in blue, converges rapidly in the usual sense. On the other hand, the contribution with $m=-s$, when summed in the usual way (purple dots), does not converge. However, by performing a Cesàro sum with order $\alpha=1$ (dashed, dark red line), or higher, e.g., $\alpha=2$ (solid red line), we can cast it into a convergent sum very accurately.
    }
    \label{fig:Sums_Convergence}
\end{figure}
\begin{equation}
    C^{(\alpha)}[z_n] = \lim_{N\to\infty} \sum_{n=0}^N \binom{N}{n}\binom{N+\alpha}{n}^{-1} z_n \, .
\end{equation}
For $\alpha=0$ this is just the usual sum, and for $\alpha = 1$ this is the limit of the average of the partial sums. 
In Fig.~\ref{fig:Sums_Convergence} we show the behavior of both sequences of partial sums, as a function of the maximum value of $\ell$ included in the sum, for positive helicity (the same is observed for the negative helicity case). As expected, the $m=s$ converges rapidly in the usual sense. The $m=-s$ case does not converge in the usual sense (this is given by the order $\alpha = 0$ Cesàro sum, shown as purple dots). By taking $\alpha=1,2$ (dashed and solid red lines), the sum becomes almost as rapidly convergent as its counterpart. Therefore we choose to separate the $m=\pm s$ contributions and sum one of them in the usual sense, and the other one in the Cesàro summation scheme with $\alpha$ sufficiently high to achieve the desired accuracy. In Fig.~\ref{fig:Error_Sums} we show the relative error in the calculation of $\mathcal{F}^+$ using two consecutive values of $\alpha$, defined as 
\begin{equation}
    \mathrm{Err}_\alpha = \frac{|\mathcal{C}^{\alpha+1}[\mathcal{F}^+_{m=-s}]-\mathcal{C}^{\alpha}[\mathcal{F}^+_{m=-s}]|}{\mathcal{F}^+} \, , 
\end{equation}
where the denominator is evaluated with $\alpha = 5$ and $\ell_{\rm max}=60$. As clearly seen in the Fig.~\ref{fig:Error_Sums}, the Cesàro summation is converging, and the error committed is below $10\%$. More importantly, we know how increasing $\ell_{\rm max}$ diminishes the error committed. A more sophisticated numerical evaluation and analysis of the convergence of the resummation would be necessary to provide highly accurate lensed waveforms in the strong field regime. This is, nevertheless, beyond the scope of this work.

A final difficulty lies at the point $\theta_{\rm OL} = 0$, where the observer, lens and source are all aligned. In that case the sequence of partial sums diverges, and a regularization scheme is necessary in order to assign a finite value to this sum. We leave exploring this for future work.

\section{Relating distances in cosmology}
\label{App:distances}

Although we consider $r_{LS}$ to only be 100$M\odot$, the distance between the observer and the lens and the source ($r_{OL}$ and $r_{OS}$) are at cosmological distances. Despite the fact that our lensing system is at a relatively low redshift ($z_S \sim 0.1$), future detectors will be able to detect similar systems at much higher redshifts, and thus the physical distances must be related to cosmology-dependent distances to account for the expansion of the Universe. 
This will provide simple relationships between the physical distances used in the strong field setup, and the cosmology-dependent distances used in the weak field setup. For convent comparison to the literature, the following expressions will not be given in geometric units.

Following \cite{Peebles:1994xt,Hogg:1999ad}, we begin by defining the comoving distance for an object at a redshift $z$ as

\begin{equation}
    D_C = \frac{c}{H_0} \int_0^z \frac{dz'}{E(z')},
\end{equation}

where $H_0$ is the Hubble constant today, and 

\begin{equation}
    E(z) = \sqrt{\Omega_M(1+z)^3 + \Omega_k(1+z)^2 + \Omega_\Lambda},
\end{equation}

with $\Omega_M, \Omega_k, \Omega_\Lambda$ being the matter, curvature, and dark energy density parameters of the Universe. 
If we assume that we are in a flat Universe (i.e. $\Omega_k = 0$), we can very easily write relate an object's redshift to the angular diameter distances commonly used in weak field lensing as 

\begin{equation}
    D_A = \frac{D_C}{1+z}.
\end{equation}

A noteworthy property of angular diameter distances for objects at redshifts $z_1$ and $z_2$ is that the difference between their angular diameter distances is not given by simply subtracting $D_A(z_2) - D_A(z_1)$. Instead, for the simple case where $\Omega_k = 0$,

\begin{equation}
    D_{A12} = \frac{1}{1+z_2} \bigg [D_{A2} - D_{A1} \bigg ].
\end{equation}

While the form of these functions becomes more complicated for open or closed universes (i.e., $\Omega_k \neq 0$), the Planck 2018 measurements of cosmological parameters were found to be consistent with a flat universe \cite{Planck:2018vyg}. 
We therefore work under this assumption to simplify calculations of cosmological effects, and direct readers to other works to see the details of nonflat Universe models (cf.~\cite{Peebles:1994xt,Hogg:1999ad}).

\end{document}